\documentclass[prb,twocolumn,showpacs,preprintnumbers,amsmath,amssymb]{revtex4}

\usepackage{amsfonts}
\usepackage{latexsym}
\usepackage{graphicx}
\usepackage{epsfig}
\usepackage{amsbsy}
\usepackage{amsmath}
\usepackage{amssymb}

\newcommand{\Eqref}[1]{Eq.~(\ref{#1})}
\newcommand{\etal}{\textit{et~al.}}
\newcommand{\nn}{\nonumber}
\newcommand{\be}{\begin{equation}}
\newcommand{\ee}{\end{equation}}

\newcommand{\alf}{{Alfv\'en~}}
\newcommand{\bear}{\begin{eqnarray}}
\newcommand{\ear}{\end{eqnarray}}
\newcommand{\Va}{V_\mathrm A}
\newcommand{\wa}{\omega_\mathrm A}
\begin{document}

\title{On the origin of solar wind. \alf~waves
induced jump of coronal temperature}

\author{T.M.~Mishonov}
\email[E-mail: ]{mishonov@phys.uni-sofia.bg}
\author{M.V.~Stoev}
\email[E-mail: ]{martin.stoev@gmail.com}
\author{Y.G.~Maneva}
\email[E-mail: ]{yanamaneva@gmail.com}

\affiliation{Department of Theoretical Physics, Faculty of Physics,\\
University of Sofia St.~Clement of Ohrid,\\
5 J. Bourchier Boulevard, Bg-1164 Sofia, Bulgaria}

\date{\today}

\begin{abstract}
Absorbtion of Alfv\'en waves is considered to be the main mechanism of
heating in the solar corona. It is concluded that the sharp increase
of the plasma temperature by two orders of magnitude is related to a
self-induced opacity with respect to Alfv\'en waves. The maximal
frequency for propagation of Alfv\'en waves is determined by the
strongly temperature dependent kinematic viscosity. In such a way
the temperature jump is due to absorption of high frequency Alfv\'en
waves in a narrow layer above the solar surface. It is calculated
that the power per unit area dissipated in this layer due to damping of
Alfv\'en waves blows up the plasma and gives birth to the
solar wind. A model short wave-length (WKB) evaluation takes into
account the $1/f^2$ frequency dependance of the transversal magnetic
field and velocity spectral densities. Such spectral densities agree
with old magnetometric data taken by Voyager 1 and recent
theoretical calculations in the framework of Langevin-Burgers MHD.
The present theory predicts existence of intensive high frequency
MHD Alfv\'en waves in the cold layer beneath the corona. It is
briefly discussed how this statement can be checked experimentally.
It is demonstrated that the magnitude of the Alfv\'en waves
generating random noise and the solar wind velocity can be expressed
only in terms of satellite experimental data. It is advocated that
investigation of properties of the solar surface as a random driver by
optical methods is an important task for future solar physics. Jets of
accretion disks are speculated as a special case of the wind from magnetized
turbulent plasma.
\end{abstract}


\maketitle

\section{Introduction}
\label{intro}
The coronal heating mechanism\cite{Aschwanden:06,Erdelyietal:03} is one of
the most perplexing longstanding unresolved problems of contemporary
physics.\cite{Mandrini00} This is the reason why even some
condensed matter theorists are involved in the revealing of this
mystery. The lack of experimental data makes distinguishing between the
various theoretical models difficult. In such a situation it is possible to
apply purely aesthetic criteria for a natural description of some well-known
facts. One such fact is that the increase of plasma temperature in the
transition region by two orders of magnitude\cite{tempgrad} occurs very
rapidly -- there is a smeared jump, which serves as a starting point for
creation of the solar wind. We wish to emphasize that
complete magneto-hydrodynamic (MHD) simulations of \alf waves
generated by a random driver also give a sudden increase of the
temperature as a function of height.\cite{Erdelyi:04} The purpose of
the current work is to give a qualitative explanation of the
observed temperature jump and to perform a model evaluation of its
order of magnitude. Finally, we will analyze some possible future
experimental observations described by the presented theoretical
scenario. For example, the model predicts existence of intensive
high frequency \alf waves beneath the coronal temperature jump.

\section{Scenario}
\label{scenario}
In order to make a proper assessment for the magnetic field spectrum
comparable with magnetometric data taken by Voyager 1\cite{Burlaga&Mish:87}
we need to switch to a frequency-dependent regime and perform a time averaging of
the Fourier transformed wave component $\mathbf B(t) \approx B_x\hat
x + B_y \hat y$ perpendicular to the constant magnetic field
$\mathbf B_0 = B_{0,z}\hat z$ along which the $z$-axis is chosen
\bear \mathbf B(t) &=& \sum_{\omega_n}\exp(-i\omega_n t)\mathbf
B_{\omega_n},\quad
\omega_n = \frac{2\pi}{\Delta t} n,\\
\nn \mathbf B_{\omega_n} &=& \frac{1}{\Delta t}\int^{\Delta t}_{0}
\exp(i\omega_n t)\mathbf B(t)\,\mathrm d t,\quad n=0,\pm1,\pm2,\dots
\ear
As our purpose is to present a model evaluation, from now on we will
consider only one of the fluctuating wave components, so the
plane magnetic field indices will be omitted. Fourier
analysis of the signal accumulated for a time interval $\Delta t$
gives us the opportunity to observe the magnetic field spectral
density $\Phi(t)$, which is tightly bound to the time averaged
square of the wave magnetic component
\be \langle B^2(t)\rangle = \frac{1}{\Delta t}\int^{\Delta t}_0
\!\!B^2(t)\,\mathrm dt =\sum_{\omega_n}{|B_{\omega_n}|}^2 =
\!\int\!\Phi(f)\,\mathrm df. \ee
The thus-defined spectral density
\bear
\nn \Phi(f) &=& \frac{1}{f_2-f_1}\sum_{\omega_n}
{|B_{\omega_n}|}^2,\quad \omega_n \in (\omega_1, \omega_2),\\
f&=&\omega/2\pi, \quad (f_2-f_1)\Delta t\gg 1 \ear
determines the magnetic field energy density $w$ and the \alf waves
energy flux $S$ as a function of the circular $\omega$ or linear
frequency $f$
\be
w = \frac{B^2(t)}{2\mu_0}, \quad S = \frac{V_\mathrm A}{2\mu_0}
\int\!\Phi(f)\,\mathrm df.
\ee
In interplanetary space the spectral density $\Phi$ is measured
in typical units nT$^2$/Hz. \alf waves energy density depends on
the integrated spectral density and the \alf speed $\Va$. In general
the integration should be taken over all possible wave frequencies
$f \in (0, \infty)$ unless the given specific power spectrum imposes
an introduction of a low or high frequency cut-off. In the suggested
model we use the natural cut-off frequency set by the condition for
waves existence $\wa\tau_\mathrm A = 1,$ when the \alf waves
frequency $\wa$ exactly equals the attenuation coefficient
${\tau_\mathrm A}^{-1}.$ Therefore in the investigation that follows we
exclude the influence of any extremely high-frequency \alf
waves that would be absorbed by the medium immediately after their
generation under the solar surface. Hence for heating in the corona
we only take into account those waves, whose characteristic life time
exceeds the inverse value of the cut-off frequency $\omega_c$
\bear
\nn \omega_\mathrm A \tau_\mathrm A &>& 1,
\quad \omega_\mathrm A = V_\mathrm A k_z, \quad 1/\tau_\mathrm A = \nu k^2\!,\\
\rho V^2_\mathrm A &=& B_{0,z}^2/\mu_0, \quad \omega_c = 2\pi f_c= V^2_\mathrm A/\nu.
\ear
Molecular kinetic theory\cite{LL10} determines the
temperature-dependence of the cut-off frequency, being a function of
the kinematic viscosity $\nu.$ A more detailed analysis shows that
the sum of kinematic and magnetic viscosity is relevant in the damping of
the Alfv\'en waves\cite{LL8,Braginskii:65}, but the magnetic viscosity is
much smaller and negligible for sufficiently high plasma temperatures.
This is because the kinamatic viscosity is $\propto T^{5/2}$ while the
resistivity is $\propto T^{-3/2}.$ For more detailed discussion see,
for example, the work by Erdelyi and Goossens\cite{ErdelyiGoosens:95}and
Ref.~\onlinecite{Goosens:03}. For low density coronal plasmas,
electromagnetic emission and thermal conduction are negligible for the energy
balance close to the temperature jump. In our model we neglect the
difference between the electron and ion temperatures which are of
the same order $T_e\approx T_p.$ However we wish to emphasize that
heating is due to ion viscosity and our consideration gives a
natural explanation why the proton temperature is higher than
electron one $T_p>T_e.$ It is because viscous friction heats
the ions.
Since we pursue just a qualitative
estimation, with a logarithmic accuracy in the final results for the
solar wind velocity presented below, we can neglect the slight
temperature variations in the protons' Coulomb logarithm $L_p$ as well
as the influence of the electron temperature dependence on the Debye
radius $a$. Precise calculations including such a dependence would
only lead to minor corrections that would not change the order of
magnitude of the final outcome
\bear
\nn \nu & =& c_\nu T_p^{5/2}, \quad c_\nu= \frac{0.4}{M^{1/2}e^4n_pL_p},
\quad L_p=\ln \frac{T_p a}{e^2}\gg1,\\
\frac{1}{a^2}&=& 4\pi
e^2\left(\frac{n_e}{T_e}+\frac{n_p}{T_p}\right)\!,
\quad e^2\equiv \frac{q_e^2}{4 \pi \varepsilon_0}.
\ear
According to valuable analysis on the experimental data obtained by
the Voyager 1 magnetometer\cite{Burlaga&Mish:87}, the magnetic field
spectral density can be approximated by a single power law. Here we
have taken into account that both magnetic and energy fluxes are
almost constant along the magnetic field lines. In such a way the
spectral parameter $\mathcal D$ on the solar surface can be
evaluated by order of magnitude if we know the satellite spectral
parameter $\mathcal D^{(\mathrm{sat})}$ and the ratio of the
constant components of the magnetic field
\be
\Phi(f) \approx \frac{\mathcal D}{f^2}, \quad
\mathcal D \simeq \frac{B_{0,z}}{B^{\,\mathrm{(sat)}}_{0,z}}\mathcal
D^{\,\mathrm{(sat)}}.
\ee
The observed power law for the energy density $\propto 1/f^2$ is
theoretically explained in the framework of Langevin-Burgers MHD
model\cite{TMYM06}. The 1D calculations for the time and noise
averaged spectral density of \alf waves generated by a white noise
random driver for the external force density, modeling the influence
of the convective stochasticity
\be \langle F(t_1,z_1)F(t_2,z_2)\rangle
=\tilde\Gamma{\rho}^2\delta(t_1-t_2)\delta(z_1-z_2) \ee
reveal the same inverse proportionality to the second power of the
\alf frequency
\be \overline E_f = \frac{\pi^2\rho \Va^2 \tilde\Gamma}{2\nu f^2 L}.
\ee
Comparison of the energy flux theoretically derived on the basis of
the Langevin-Burgers approach applied for modeling the role of the
turbulence for generation of \alf waves with the experimentally
observed energy flux can give us a reliable assessment for the
Burgers parameter $\tilde \Gamma$
\be
S = L\!\int\!\Va\overline E_f\, \mathrm df \!\!=\!\! \int
\frac{\pi^2\rho\Va^2\tilde\Gamma}{2\nu f^2}\,\mathrm df \!=\!\int
\!\!\frac{\Va\mathcal D}{2\mu_0f^2}\,\mathrm df. \ee
Thus, if we consider Burgers approach as adequate for a turbulence
model description, we can extract information for the turbulence
spectrum in the photosphere, at the footpoints of the magnetic field
lines
\be
\label{Burgersparam}
\tilde\Gamma = \mathcal
D\nu/\pi^2\mu_0\rho\Va.
\ee
According to a recently proposed
scenario\cite{TMYM06,MishonovStoev:07} in the spirit of earlier
ideas for wave heating\cite{Schatzman,Ionson:78} \alf waves serve as
mediators, carriers of energy from the turbulent photosphere to the
hot solar atmosphere, where in a small region the high-frequency
waves attenuate intensively and heat the corona.Ref.~\cite{TMYM06}
treats the emission, while Ref.~\cite{MishonovStoev:07} considers the bulk
absorption of \alf waves. In this work we present an evaluation for the
absorbed in the transition zone energy flux, whose strong temperature
dependence is determined by the temperature dependence of the cut-off
frequency $f_c$ and naturally results in a sharp temperature
jump\cite{MishonovStoev:07}
\be
f_c = \frac{V^2_\mathrm A}{2\pi c_\nu T^{5/2}}.
\ee
The absorbed energy flux is taken from Voyager 1 magnetometric data analysis,
but it may also be obtained by a thermodynamical approach. If the
comparatively small effects associated with radiative losses and
compression are neglected, the absorption rate will be related only
to the plasma internal energy density $\varepsilon$ and the
conducted work
\bear S_\mathrm{Abs}\!&=&\!V_\mathrm A
\int_{f_c}^{\infty}\!\!\frac{\mathcal D}{2\mu_0f^2}
\,\mathrm df = \frac{\Va \mathcal D}{2\mu_0 f_c}=
\frac{\pi c_\nu \mathcal D}{\mu_0V_\mathrm A}T_p^{5/2}\\
\nn &\approx & \varepsilon v + pv, \quad \varepsilon = \frac{3}{2}p,
\quad p= n_eT_e + n_pT_p, \ear
where $p$ and $v$ are respectively the plasma pressure and velocity.
In such a way we can derive the approximate rate for the velocity of
the solar wind, driven by sharp coronal temperature increase due to
absorption of intensive high frequency \alf waves, for which the
transition region plasma is opaque
\be
\label{windspeed}
v_\mathrm{wind} \simeq \frac{0.08\,\pi
\mathcal DT_p^{3/2}}{\mu_0 \Va M^{1/2}e^4n^2_pL_p}, \quad T_p\sim
T_e. \ee
As the shear viscous friction heats the heavy particles,
\cite{Ofmanetal:94,ErdelyiGoosens:95} the proton temperature $T_p$ is
significantly higher than that of the electrons $T_e,$ however, for an
order of magnitude evaluation here we suppose the electron temperature
$T_e$ to be similar to that of the protons $T_p.$

\section{Discussion and conclusions}
\label{concl}
With a logarithmic accuracy we have derived an
explicit formula for the initial velocity of the solar wind. This
formula \Eqref{windspeed} is completely based on experimentally
accessible parameters and can be easily rejected if it gives more
than 3 orders of magnitude difference. But, if this model remains valid,
let us briefly discuss what has to be done as a future
perspective in order to finally solve the perplexing longstanding
mystery for the origin of coronal heating and solar wind. First of
all, numerical simulations on MHD with a white noise random
driver\cite{Erdelyi:04} (Langevin-Burgers MHD) have to be repeated to
reproduce a $\delta$-like maximum of the energy dissipation density
at the transition region. In other words all theoretical models
qualitatively explaining the Voyager 1 data for the frequency
dependent spectral density of the magnetic field have to be compared
with random driver computer simulations of coronal heating.
Confirmation of a narrow maximum of volume heating power due to
self-induced plasma opacity is a routine task for further numerical
investigations. The numerical analysis could improve the present
analytical evaluation incorporating, for example, the reflection of
\alf waves by the jump in plasma density.

Here we wish to insert a short historical remark. Stochastic
mechanics in general was introduced by Langevin\cite{Langevin:08}
in 1905 to explain the Brownian motion. Later, Burgers\cite{Burgers:48} in 1948
introduced the white noise random driver in the hydrodynamics of
turbulence. Much later, in 1995, Polyakov\cite{Polyakov:95} derived
Kolmogorov power laws\cite{Kolmogorov:41} using Langevin-Burgers
approach, but his research remained unobserved in astrophysics, not to
speak about heliophysics. That is why using random number generator
in MHD simulations is simply called \emph{random driver}, but the magnitude
of the noise is almost never evaluated by comparison to real experimental data.

In the recent work we have only evaluated the area under the sharp
maximum of dissipation and now it is time to perform state of the
art calculations with a realistic random force, whose spectral
density corresponds to the data taken by the satellites'
magnetometers. For example, the turbulence spectrum can be treated
with the based on the Langevin-Burgers model assessment
\Eqref{Burgersparam}. We propose that the physical mechanism for
the origin of solar wind and coronal heating is already revealed and
it is only a matter of honest numerical work to create a coherent
picture. The results for the temperature jump and dissipation
maximum have to be compared with the observations and other
theoretical scenarios such as magnetic
reconnections,\cite{Priest&Longcope&Heyvaerts:05} for example. Each
model unable to reproduce a sharp temperature increase in height has
to be assigned to the waste basket. The same can be said for
the models based on Ohmic heating which predict an electron temperature
higher than that of the protons. Though the latter statement seems to be
true for the X-ray bright points, it is definitely not valid for the
corona as a whole. We are unaware, for example, how magnetic
reconnection theory can explain why the proton temperature is higher
than that of the electrons $T_p>T_e.$ Analogously we have not found
any reference explaining how turbulent cascade can lead to a
very sharp increase of the coronal temperature.

Second, in order to evaluate the properties of the solar surface as a
random driver, all the old data from Voyager~1 has to be meticulously
analyzed and a detailed investigation by the forthcoming Solar
Orbiter mission has to be planned.

Third, the Sun is a unique system for investigation of convective
turbulence. It will be very interesting to compare the satellites'
data for the magnetic field spectral density with results based on
theoretical modeling of turbulence. The maturity of solar physics
can stimulate significant development of the achievements in
contemporary turbulence research. For instance, the $1/f^2$ power law by
Burlaga and Mish\cite{Burlaga&Mish:87} corresponds to
one-dimensional $(k_x, k_y=0)$ propagation of \alf waves in the
framework of Langevin MHD,\cite{TMYM06} as the whole noise is
created by the random motion of the funnel foot-points.

Next, absorbption of \alf waves could be an important mechanism
in many cases of space plasmas including accretion disks as well.
MHD waves in accretion disk amplification by shear flow is the main
mechanism of transformation of gravitational energy into heat; for
references see the preprint.\cite{TMYMTH:05} The magnetized wind
from an accretion disk should follow the magnetic force lines and
plasma concentrated above the disk center will create an accumulative
jet perpendicular to the disk. In such a way disk jets could be recognized
as a wind coming from magnetized turbulent plasma.

Lastly, now we operate with a realistic 3D model for the
distribution of the magnetic field from the solar surface to the
satellite. The perturbation of magnetic field lines serves as a
string of a harp to deliver the information about the solar
turbulence from the photosphere to the magnetometer. Owing to
the propagation of \alf waves we can ``listen'' to the sounds of the
great solar symphony. Due to absorption of the high frequency modes,
however, at the transition region where the temperature jump occurs
we are able to hear only the basses, whereas for the ultra-violin
band we remain absolutely deaf. A current problem which deserves
to be put on the agenda is to observe the powerful high frequency
\alf modes \emph{under} the corona.\cite{Nakariakov:05} A kamikaze
satellite could give some very important information, but for
systematic research we need to learn how to extract the behavior of
the solar surface as a random driver using optical data. We conclude
that the first important step in this direction is to establish
correspondence between satellite magnetometric data and Doppler
shift spectra for some bright events on the solar surface. Only after a
proper incorporation of these ingredients we can conclude that our
understanding of heating mechanism of the solar corona is complete and
we have disclosed a very important case of heating of space plasmas.

\acknowledgments

Support and fruitful discussions with D.~Damianov, A.~Rogava,
T.~Zakarashvili, and R.~Erd\'elyi are highly appreciated.



\begin{thebibliography}{99}
%
\bibitem{Aschwanden:06}
M.~J.~Aschwanden, \textit{Physics of the Solar Corona: an Introduction with
Problems and Solutions}, 2nd ed., Springer Berlin (2006).
%
\bibitem{Erdelyietal:03}
R.~Erd\'elyi \textit{et.~al.} (eds) \textit{Turbulence, Waves and
Instabilities in the Solar Plasma} in \textit{NATO Science Series II:
Mathematics, Physics and Chemistry}, Vol.~\textbf{124}, Kluwer,
pp.1-388 (2003) [ISBN: 1-4020-1658-1].
%
\bibitem{Mandrini00} C.~H.~Mandrini, P.~D\'emoulin and J.~A.~Klimchuk,
``Magnetic field and plasma scaling laws: their implications for
coronal heating models'', ApJ, \textbf{530} 999-1015 (2000).
%
\bibitem{tempgrad} M.B.~Larson in Keneth R.~Lang,
\textit{Sun Earth and Sky} (Springer-Verlag, Berlin, 1995);
http://solar.physics.montana.edu/YPOP/Spotlight/
SunInfo/transreg.html.
%
\bibitem{Erdelyi:04} R.~Erd\'elyi and S.P.~James,
``Can ion-damping help to form spicules? II Random driver'', A\&A,
\textbf{427}, 1055-1064 (2004), Figs.~8 and 10.
%
\bibitem{Burlaga&Mish:87} L.F.~Burlaga and W.H.~Mish,
J. Gephys. Res., \textbf{92}, 1261 (1987);
E.~Marsch, ``MHD Turbulence in the Solar Wind''
in \emph{Physics and Chemistry in Space --
Space and Solar Physics}, Vol.~21,
Series Editors: M.C.E.~Huber \etal,
\textit{Physics of the Inner Heliosphere},
Vol.~2, Editors: R.~Schwenn and E.~Marsch, Springer-Verlag Berlin
(1991); Fig.~10.4.
%
\bibitem{LL10} L.~Landau and E.~Lifschitz,
\textit{Course of Theoretical Physics}, Vol.~10 \textit{Physical
Kinetics}, (Pergamon, New York, 1981), Chap.~IV, Sec.~43,
Eqs.~(43.8-10).
%
\bibitem{LL8} L.~Landau and E.~Lifschitz,
\textit{Course of Theoretical Physics}, Vol.~8
\textit{Electrodynamics of Continuous Media}
(Pergamon, New York, 1993),
Chap.~VIII~ Sec.~69.
%
\bibitem{Braginskii:65}
S.~I.~Braginskii, Rev. Plasma Phys., \textbf{1}, 205 (1965).
%
\bibitem{ErdelyiGoosens:95} R.~Erd\'elyi and M.~Goossens,
``Resonant absorption of Alfven waves in coronal loops in
visco-resistive MHD'', A\&A, \textbf{294}, pp. 575-586 (1995).
%
\bibitem{Goosens:03}M.~Goossens,
\textit{An introduction to plasma astrophysics
and magnetohydrodynamics}, Kluwer London, p.~108 (2003);
see Chap.~4, eq.~(4.38) and explanations thereafter.
%

\bibitem{TMYM06} T.~Mishonov and Y.~Maneva,
``Burgulence and Alfv\'en Waves Heating Mechanism of Solar Corona'',
astro-ph/0609609.
%
\bibitem{MishonovStoev:07} T.~Mishonov, M.~Stoev and Maneva,
``Theory of heating of hot magnetized plasma by ALfv\'en waves.
Application for solar corona'', astro-ph/0701554.
%
\bibitem{Ofmanetal:94}
L.~Ofman, J.~M.~Davila and R.~S.~Steinolfson,
``Coronal heating by the resonant absorption of Alfven waves:
The effect of viscous stress tensor'', ApJ \textbf{421}, p. 360-371
(1994).
%
\bibitem{Schatzman} E.~Schatzman,
Annales d'Astrophysique, \textbf{12}, 203 (1949); M.~Schwarzschild,
Astrophysical Journal, \textbf{107}, 1 (1948).
%
\bibitem{Ionson:78} J.A.~Ionson, ``Resonant absorption of Alfv\'enic
surface waves and the heating of solar coronal loops'', ApJ,
\textbf{226}, 650--673 (1978).
%
\bibitem{Langevin:08} P.~Langevin, Comptes. Rendues Ac. Sci. Paris,
\textbf{146}, 530 (1908).
%
\bibitem{Burgers:48} J.M. Burgers, Adv. in Appl. Mech.
 \textbf{1}, 171 (1948).
%
\bibitem{Polyakov:95}
A.M.~Polyakov, ``Turbulence without pressure'', Phys. Rev.~E
\textbf{52}, 6183–6188 (1995); hep-th/9506189.
%
\bibitem{Kolmogorov:41} A.N.~Kolmogorov,
 ``The local structute of turbulence in incompressible
 viscous fluid for very large Reynolds numbers,''
 Comptes Rendue (Doklady) Acad. Sci. URSS (N.S.)
 \textbf{30}, 301--305 (1941).
%
\bibitem{Priest&Longcope&Heyvaerts:05} E. R.~Priest, D. W.~Longcope and
J.~Heyvaerts, ``Coronal Heating at Separators and Separatrices'', ApJ,
\textbf{624}, pp. 1057-1071 (2005).
%
\bibitem{TMYMTH:05} T.M.~Mishonov, Y.G.~Maneva, T.S.~Hristov, \textit{``On
the theory of MHD waves in a shear flow of a magnetized turbulent plasma''},
astro-ph/0507696.
%
\bibitem{Nakariakov:05}
L.~Ofman, J.M.~Davila, V.M.~Nakariakov, \textit{et al.},
``High-frequency Alfven waves in multi-ion coronal plasma:
Observational implications'', Journal Of Geophysical Research
\textbf{110}  (A9), A09102, 0148--0227 (2005).



\end{thebibliography}
\end{document}